\documentclass[aps,pre,twocolumn, amsmath,amssymb,showpacs,superscriptaddress,floatfix]{revtex4-1}
\usepackage{graphicx}
\usepackage{slashbox}
\usepackage{color}

\begin{document}
\title{Attractive and repulsive polymer-mediated forces between scale-free surfaces}

\author{Yacov Kantor}
\email{Email: kantor@post.tau.ac.il}
\affiliation{Raymond and Beverly Sackler School of Physics and Astronomy,
 Tel Aviv University, Tel Aviv 69978, Israel}
\author{Mehran Kardar}
\affiliation{Massachusetts Institute of Technology, Department of
  Physics, Cambridge, Massachusetts 02139, USA}

\date{\today}


\begin{abstract}
We consider forces acting on objects immersed in, or attached to, long fluctuating polymers.
The confinement of the polymer by the obstacles results in polymer-mediated forces
that can be repulsive (due to loss of entropy) or attractive
(if some or all surfaces are covered by adsorbing layers).
The strength and sign of the force in general depends on the detailed shape and adsorption
properties of the obstacles, but assumes simple universal forms if characteristic length scales
associated with the objects are large.
This occurs for scale-free shapes (such as a flat plate, straight wire, or cone),
when the polymer is repelled by the obstacles, or is marginally attracted to it
(close to the depinning transition where the absorption length is infinite).
In such cases, the separation $h$ between obstacles is the only relevant macroscopic
length scale, and the polymer mediated force equals ${\cal A} \, k_{B}T/h$, where $T$
is temperature. The amplitude ${\cal A}$ is akin to a critical exponent,
depending only on geometry and universality of the polymer system.
The value of ${\cal A}$, which we compute for simple geometries and ideal polymers,
can be positive or negative.
Remarkably, we find ${\cal A}=0$ for ideal polymers at the adsorption transition point,
irrespective of shapes of the obstacles, i.e. at this special point there is no polymer-mediated
force between obstacles (scale-free or not).
\end{abstract}
\maketitle

\section{Introduction}

A prototype of soft matter, polymers are long flexible chains that can
fluctuate (whether within a cell or in a solution) between a large number
of configurations. The presence of hard boundaries or obstacles modifies
the number and weight of allowed configurations, in turn resulting in
{\it polymer mediated forces} between the obstacles.
A well known example
is the {\it depletion force} of polyethylene glycol (PEG) which acts to
bundle filaments~\cite{Dogic16}. However, whereas the relevant length scale
for depletion force is the overall size of the polymer $R$, here we focus on
polymer-mediated forces on separations $h\ll R$. The internal structure of a
polymer is a self-similar fractal, spanning a wide range of scales from $R$
to a microscopic monomer size $a$. To compute forces between obstacles
embedded in or attached to the polymer, we need to compute modifications to
the free energy due to the objects. This is in principle a complex task
involving the shapes of the objects, and details of their interactions with
the polymer. We demonstrate that this task is considerably eased in
certain cases, yielding simple universal expressions for the force.

Technological progress in manipulation of single molecules~\cite{bustamante03,kellermayer,neuman,deniz,neuman08}, using probes such as atomic force microscopes (AFMs)~\cite{fisher}, microneedles~\cite{kishino},
optical~\cite{neuman2004,hormeno} and magnetic~\cite{gosse} tweezers,
makes it possible to measure forces exerted by polymers with high precision.
The central motivation of these experiments is to unravel specific information about
shapes, bindings, and interactions of biological molecules from force-displacement curves.
For the important class of intrinsically unstructured proteins~\cite{Dyson05},
entropic forces, such as those considered in the paper, are likely to play an important role.

In previous work we considered polymers confined by impenetrable obstacles
of scale invariant shape, such as a polymer attached to the tip of a conical
probe approaching a flat surface~\cite{MKK_EPL96,MKK_PRE86,HK_PRE89}. The
reduction in the number of configurations of the polymer leads to a
{\it repulsive} entropic force, which we showed to depend on the (tip
to surface) separation $h$ and the temperature $T$ as $F={\cal A}\, k_{B}T/h$.
The ``universal" amplitude ${\cal A}$ only encodes basic geometrical properties,
and gross features of the polymer. For such ``repulsive" surfaces, the
amplitude $\cal A$ is positive. By considering both repulsive and attractive
surfaces (as well as by expanding the types of polymers considered), here we
demonstrate that attractive surfaces may indeed lead to polymer-mediated
attraction with negative ${\cal A}$.

The interaction of a polymer with a surface can be changed from {\it repulsive}
to {\it attractive}, e.g. by changing temperature or solvent quality. The
competition between energetic attraction and entropic repulsion typically leads
to a temperature dependent absorbed layer size, introducing another length into
the problem. This length scale diverges at a continuous adsorption transition
point introducing a scale-free boundary condition which is distinct from
the repulsive surfaces studied previously.

In this paper we expand our formalism~\cite{MKK_EPL96,MKK_PRE86,HK_PRE89,AK_PRE91}
from the treatment of purely repulsive surfaces to adsorbing surfaces, and to
mixed repulsive/adsorbing surface combinations.
In Section~\ref{sec:SurfaceExponents} we begin  examining  several
polymer types near repulsive or adsorbing {\em flat} surfaces, and show that
the size and sign of the force between a polymer and a surface depends both on the
polymer type and the surface type. In Section~\ref{sec:forces} we demonstrate
that under certain circumstances the polymer-mediated forces between scale-free
surfaces have a universal coefficient, independent of  minute details of the
polymers. The calculation of force induced by ideal polymers, taken up in
Section~\ref{sec:ideal}, can be reduced to the solution of a diffusion problem
with either absorbing or reflecting boundary conditions. A particularly
interesting result is that when all embedded surfaces are at the adsorption
transition point, the polymer
mediated force is identically zero, independent of shape and geometry. In
Sections~\ref{sec:mixed2D} and \ref{sec:mixed3D} we consider a number of examples
of mixed repulsive/attractive geometries and demonstrate the ability to modify
the force amplitude by changing the surface geometries and types. Finally, under
Discussion we consider possible generalizations of ideal polymer results
to other polymer types.

\section{Polymers near attractive or repulsive flat surfaces}\label{sec:SurfaceExponents}

\begin{table}
\vspace{.4cm}
\begin{tabular}{| l ||  c | c | c | c|}
\hline
\backslashbox[30mm]{polymer type}{exponent} & $\nu$ & $\gamma$ & $\gamma_1$ & $\gamma_a$\\ \hline \hline
ideal polymer at any $d$ & 1/2 &  1 & 1/2 & 1\\ \hline
SA polymer at $d=2$ & 3/4 &  43/32  & 61/64 & 93/64 \\ \hline
SA polymer at $d=3$  & 0.588 & 1.157 & 0.697 & 1.304 \\ \hline
$\theta$-polymer at $d=2$  & 4/7  & 8/7 & 4/7 & 8/7 \\ \hline
\end{tabular}
\caption{\label{tab:exponents} Exact (simple fractions) and approximate (decimal fractions)
values of exponents of ideal, self-avoiding (SA) and $\theta$ polymers in free space
and close to repulsive or critical attractive {\em flat} surfaces~\cite{JansevanRensburg15a}.
(The exponents are defined in the text.)}
\end{table}

Polymers may exist in different phases, with distinct universal
characteristics~\cite{degennesSC}. At  high temperatures in a good
solvent polymers expand to maximize the number of available configurations.
Ignoring all interactions between monomers, except those imposing its
connectivity leads to configurations resembling a random walk;
such configurations will be denoted as {\it ideal} polymers.
However, it is unrealistic to ignore the exclusion of monomers from
occupying the same volume in space, and the resulting configurations
(which are more swollen than ideal random walks) are designated as
{\em self-avoiding} polymers. When the  quality of a solvent is
reduced, the tendency of monomers to aggregate is akin to an effective
short-range attraction which eventually collapses the polymer to a
globule of finite density. The transition between  good and bad
solvent regimes occurs at the so-called $\theta$-point, with the resulting
configurations labeled as $\theta$-polymers.

Ideal, self-avoiding and $\theta$ polymers are the three polymer types considered
in this work, all characterized by (albeit distinct) {\em universal}
scale-invariant properties.
For example, they are characterized by a fractal dimension $d_f=1/\nu$,
such that the typical separation between monomers $i$ and $j$ along the chain
scales as $|i-j|^\nu$; the overall polymer size $R$ (such as the mean radius of gyration, or
the end-to-end distance) grows with the number of monomers $N$ as $R=aN^\nu$,
where $a$ is some microscopic length, of the order of monomer size or persistence length.
The exponent $\nu$ depends only on polymer type, but not on any microscopic details.
It ranges from 1/2 to 3/4 depending on space dimension $d$ and the polymer type,
as listed in Table~\ref{tab:exponents}. This universality enables the
frequent use of simple lattice models to study real polymers. For example,
ideal and self-avoiding polymers can be represented by random walks and self-avoiding walks on
lattices, respectively, while $\theta$ polymers may be represented as self-avoiding
walks on a lattice
with added attractive interaction between monomers on adjacent lattice sites.
(The attractive interaction must then be tuned to exactly match the boundary
between  good and  bad solvents.)

The partition function of polymer types described above is in part
universal~\cite{degennesSC}. It depends on the number of monomers as
\begin{equation}\label{Eq:gammadef}
{\cal Z}= b \, z^N  N^{\gamma-1},
\end{equation}
where  $b$ and $z$ depend on microscopic properties of the polymer,
while the power-law exponent $\gamma$ depends only on geometry
and polymer type. Thus, the
leading {\em extensive} part of the free energy of a single polymer
\begin{equation}\label{Eq:Helmholtz}
{\cal F}=-k_BT\ln{\cal Z}=-k_BTN\ln z-k_BT(\gamma-1)\ln N+\dots
\end{equation}
is model dependent, while the coefficient of the {\em subleading} $\ln N$
is universal. Nevertheless, we shall see that this subleading term
plays an important role in polymer-mediated forces. In self-avoiding and ideal polymers,
the potential energy plays a minor role. In lattice models it is completely
absent, and $\cal Z$ coincides with the total number of configurations $\cal N$,
while $z$ is the lattice coordination number for random walks, or the effective
coordination number for self-avoiding walks.
The free energy is then obtained from the entropy $\cal S$
as ${\cal F}=-T{\cal S}=-k_BT\ln{\cal N}$.

If one end of a polymer is attached to an infinite impenetrable flat surface in $d=3$,
or to an infinite repulsive line in $d=2$, then it will be excluded from half of
the space. Nevertheless, the metric exponent $\nu$ remains unmodified, although the
prefactor $a$ in the power law $R=aN^\nu$ does change. The number of
available configurations, and hence the partition function, is reduced to
${\cal Z}_1=b_1z^NN^{\gamma_1-1}$.
Note that the factor $z$ related to the extensive part of the free energy
is unchanged, with the reduction in states captured through the
exponent $\gamma_1<\gamma$ (see Table~\ref{tab:exponents}).
The change in free energy
\begin{equation}\label{Eq:DeltaF1}
\Delta{\cal F}_1\equiv{\cal F}_1-{\cal F}=k_BT(\gamma-\gamma_1)\ln N
\end{equation}
is positive, i.e. the polymer is repelled by the wall, or, a force
towards the wall needs to be applied to bring the polymer from infinity
to the wall.

If the repulsive surface described above is covered by an attractive
layer, then a polymer attached by one end to the surface may decrease
its energy by frequently visiting the surface. In discrete models we
may simply assign an extra (Boltzmann) weight $q={\rm e}^{-V/k_BT}$,
where $V<0$ is the potential at the attractive layer, for each point
visited at the boundary.
The reduction in entropy of the polymer in this {\em absorbed} state
is compensated by a bigger gain in energy.
At high temperatures (or for weakened attractive potential) the entropy
wins and the polymer depins from the attractive layer.
The free energy per monomer in the absorbed state is
lower than that of the free polymer due to the gain in absorption energy,
and can be cast as  $-k_BT\ln z_a(T)$ with $z_a(T)>z$.
If one end of the polymer is held at some moderate distance $h$ from the surface,
then a typical configuration will consist of an ``equilibrium bulk" attached to the surface,
and a strongly stretched tail going from the surface to the point where it is held (with a
force of order of $-(k_BT/a)\ln [z_a(T)/z]$).
For some computations, it is more convenient to consider a slightly different situation
where the polymer is anchored to the surface and pulled away by application of a
force~\cite{Skvortsov12,JensevanRensburg13,Orlandini16}.
In this situation, the behavior of the polymer is different if we
control the distance $h$ {\em versus}  the pulling force applied to its end,
akin to controlling density {\em versus}
pressure at a first order liquid-gas transition~\cite{Skvortsov12}.

The transition from adsorbed to desorbed states occurs at a critical
(depinning) temperature $T_a$~\cite{Eisenriegler82,Binder83,debell,Livne88,Meirovitch88,Meirovitch93,EisenrieglerBook93,Vrbova98,Rychlewski11},
where $z_a(T_a)=z$.
Exactly at $T_a$, the partition function of any of the polymer types
mentioned~\cite{JansevanRensburg15a} above again has a simple form
${\cal Z}_a=b_az^NN^{\gamma_a-1}$. Since almost all monomers are away from
the boundary (the fraction of contacts with the boundary increases slower
than $N$), the dominant factor of $z$ remains unchanged. The relation
between the exponent $\gamma_a$ and the free-space $\gamma$ is not
obvious, since the presence of the surface decreases the number of available
configuration, which tends to decrease $\gamma$, but also decreases
the energy, which tends to increase $\gamma$.  By comparing $\gamma$
with $\gamma_a$ in the Table \ref{tab:exponents}, we see that for
self-avoiding polymers $\gamma_a>\gamma$, while for ideal and $\theta$ polymers
$\gamma_a=\gamma$. This means that
\begin{equation}\label{Eq:DeltaFa}
\Delta{\cal F}_a\equiv{\cal F}_a-{\cal F}=k_BT(\gamma-\gamma_a)\ln N\,,
\end{equation}
is either negative, i.e., the polymer is attracted by the wall, or vanishes,
which makes the wall ``invisible" to the polymer that is brought into
its vicinity. When $T$ is not at the adsorption transition point
we may expect deviations from the above relations and various crossover
effects. However, as long as the
correlation length $\xi$ characterizing the transition~\cite{debell}
exceeds the polymer size, we may treat
the system as if it is at $T_a$. In the remainder of this article
we will always assume that the attractive surfaces are at adsorption
transition point without explicitly mentioning this condition.

\section{Polymer-mediated forces between scale-free surfaces}\label{sec:forces}

\begin{figure}
\includegraphics[width=8cm]{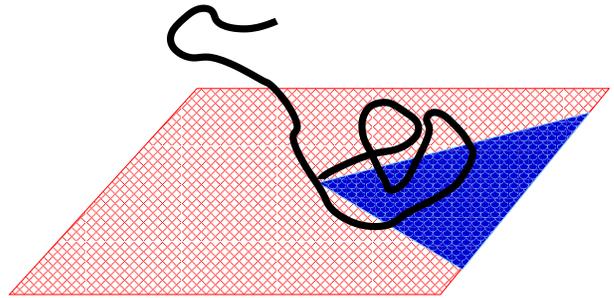}
\caption{
\label{fig:DecoPlane}
(Color online) A repulsive surface decorated by a sector with central angle
$2\alpha$ covered by an adsorbing layer. The polymer is attached to
the apex of the sector.
}
\end{figure}

The results in the previous section relied on the observation that
the partition function of a polymer in free space or near a planar surface
(either repulsive or at adsorption transition point) has  the form in
Eq.~\eqref{Eq:gammadef}. This form is a consequence of the fact that the
geometries of free space or infinite plane do not posses a characteristic
geometrical length scale, i.e., the relevant space is invariant under the
coordinate transformation ${\bf r}\to\lambda{\bf r}$.
Similarly, the polymer/surface interactions do not introduce a length
scale when they are either repulsive or attractive at adsorption transition
point. The same conclusion [hence Eq.~\eqref{Eq:gammadef}] applies to a host
of other {\em scale-free} shapes such a semi-infinite plane, a sector
of a two-dimensional plane in $d=3$, a semi-infinite line, a cone of any cross section,
 a wedge, or any combinations of such shapes, such as a cone touching a plane.
Scale invariance in most such geometries is with respect to a
``center" location, such as the apex of a cone or the terminal point of semi-infinite line.
We assume that in such cases an attached polymer is anchored to the
``center" point to avoid introducing a new length scale.
The partition function of a polymer attached to the central point of any scale-free
shape will be described by Eq.~\eqref{Eq:gammadef}, with an exponent
$\gamma$ that depends on $d$, the polymer phase, surface adsorption
(repulsive or attractive), and on geometric features characterizing the shape,
such as the apex angle $\Theta$ of the cone~\cite{MKK_EPL96,MKK_PRE86,HK_PRE89}, or the tilt angle of the cone touching a plane~\cite{AK_PRE91}.
Furthermore, we can mix surface types, by, say, attaching a cone
with attractive cover to a repulsive plane. In fact we can have a scale-free situation
when a single surface mixes repulsive and attractive regions: E.g.For example,
consider a repulsive plane on which a sector has been covered by an
attractive layer as in Fig.~\ref{fig:DecoPlane} (with the polymer attached
to the sector apex).

\begin{figure}
\includegraphics[width=7cm]{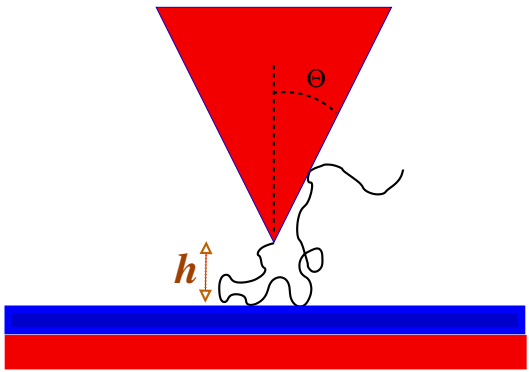}
\caption{A polymer (in $d=3$) attached to the tip of a repulsive solid  cone with apex
angle $\Theta$ approaches a plane covered by an attractive (blue)
layer. The distance between the plane and the tip of the cone is $h$. (This picture
can  also be interpreted as a two-dimensional wedge approaching a line.)
\label{fig:def_geometry}}
\end{figure}

Starting from the polymer partition function in scale-free geometries, we
can compute polymer-mediated forces between such surfaces.
As an example consider a repulsive cone, with a polymer attached  to its tip,
approaching, say, an attractive plane, as depicted in Fig.~\ref{fig:def_geometry}.
When the distance $h$ between the cone and the plane is significantly shorter
that the polymer size $R$, but larger than the microscopic scale $a$,
$h$ is the only relevant length scale, while $k_BT$
is the only relevant energy scale. In such a case, the force $F$
transmitted by the polymer between the surfaces is constrained to be the only
dimensionally correct combination
\begin{equation}
\label{Eq:force}
F={\cal A}\,\frac{k_BT}{h}\ .
\end{equation}
The dimensionless prefactor $\cal A$ (the ``force amplitude") can be
positive or negative corresponding to polymer-mediated
repulsion or attraction between the objects. (This also follows from
various polymer scaling forms~\cite{Eisenriegler82,Duplantier86,Row11}.)
Note that the form of the force (and independence of $R$) is a consequence
of the objects having a single point of closest approach.
Most of the polymer-surface interactions appear in the neighborhood of this point,
while the remote tail of the polymer is not much influenced by the constriction.
Equation~\eqref{Eq:force} fails in truly confining geometries; e.g.,
if confined between parallel planes a distance $H$ apart, the polymer has
nowhere to escape and the total polymer-mediated force can be viewed as a sum
of forces exerted by Pincus--de Gennes blobs~\cite{Pincus76,deGennes1976}
whose size depends on $H$, while their number is proportional to
$N$~\cite{degennesSC}, leading to a force proportional $N$.

\begin{figure}
\null\vskip 1cm
\includegraphics[width=8cm]{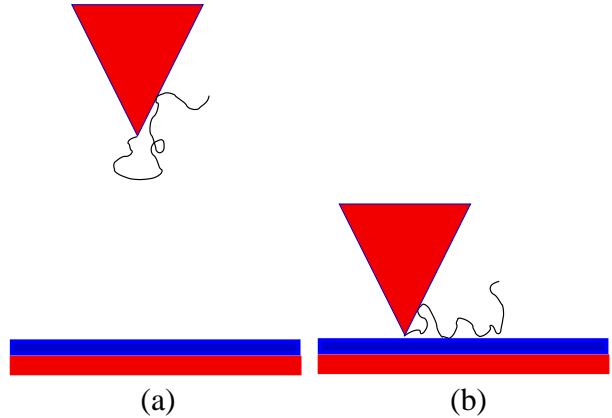}
\caption{
\label{fig:extremes}
(Color online) As the cone or wedge depicted in Fig.~\ref{fig:def_geometry}
approaches the surface, it moves between two extreme geometries: (a) When $h\gg R$
we have a scale-free geometry of a polymer attached to the cone or wedge.
(b) On touching the surface (or at a microscopic distance
$a$ from it), we arrive at the scale-free geometry of the cone+plane.
}
\end{figure}
Equation~\eqref{Eq:force} is valid only for $h$ ranging between $a$ and $R$.
When $h$ decreases and approaches the microscopic size $a$, the force
saturates at order ${\cal A}k_BT/a$,
while for $h$ exceeding $R$ it rapidly drops to 0.
Thus, the work
that the external force needs to perform to bring the surfaces from far away
 to a microscopic distance is
\begin{equation}
\label{Eq:work}
W=\!\int_a^{R}\!\!{\rm d}h~{\cal A}\frac{k_BT}{h}\ =\!{\cal A}k_BT\ln\frac{R}{a}
 ={\cal A}\,\nu k_BT\ln N.
\end{equation}
(The slight uncertainty in the integration limits is not important since it
only affects an additive constant to a term that diverges as $\ln N$.)
The same work can also be computed from the change in free energies between the final
and initial states. Both the initial and final states are scale-free as
depicted in Fig.~\ref{fig:extremes}: Far away only the cone needs to be taken
into account, while at the point where the cone touches the plane, we again
have a scale-free situation. Therefore, the partition functions in these two
extremes will have the form of Eq.~\eqref{Eq:gammadef}, but with exponents
$\gamma_{\rm far}$ and $\gamma_{\rm near}$ corresponding to the two limiting
geometries, with appropriate polymer and surface types in dimension $d$.
As in Eqs.~\eqref{Eq:DeltaF1} and \eqref{Eq:DeltaFa} the free energy difference is
\begin{equation}
\label{Eq:DF}
\Delta {\cal F}\equiv{\cal F}_{\rm near}-{\cal F}_{\rm far}
=k_BT(\gamma_{\rm far}-\gamma_{\rm near})\ln N  .
\end{equation}
By equating this $\Delta {\cal F}$ with the work $W$ in Eq.~\eqref{Eq:work} we find
\begin{equation}
\label{Eq:A}
{\cal A}=\frac{\gamma_{\rm far}-\gamma_{\rm near}}{\nu}=\eta_{\rm near}-\eta_{\rm far}\, .
\end{equation}
In the final step we employed the exponent identity
\begin{equation}
\label{Eq:gammanu}
\gamma=(2-\eta)\nu\,,
\end{equation}
to relate the exponent $\gamma$ to the exponent $\eta$ characterizing the
anomalous decay of density correlations (as~$1/r^{d-2+\eta}$).
Equation~\eqref{Eq:A} indicates that the force amplitude $\cal A$ is a universal
quantity akin to critical exponents.
In the trivial case, when the polymer, held by a very small probe (point),
is moved towards a plane, $\gamma_{\rm far}$ coincides with $\gamma$ of
the free space, while $\gamma_{\rm near}$ is $\gamma_1$ or $\gamma_a$ for
repulsive or attractive surfaces, respectively. For example, for a self-avoiding polymer
in $d=3$ approaching an attracting surface, Eq.~\eqref{Eq:A} with the exponents from
Table~\ref{tab:exponents} leads to  ${\cal A}\approx -0.25$.

The case of purely repulsive boundaries was considered previously
for both ideal and self-avoiding polymers.  The latter
required either numerical simulations or resorting to expansions in
$\epsilon=4-d$~\cite{SZKK_PRL94,MKK_EPL96,MKK_PRE86} to compute
the relevant exponents, while ideal polymers could be treated
analytically for simple (highly symmetric) geometries~\cite{MKK_EPL96,MKK_PRE86,HK_PRE89},
only requiring simple numerical solutions of diffusion equations for less symmetric scale-free geometries~\cite{AK_PRE91}.

\section{Ideal polymers near repulsive and attractive surfaces}\label{sec:ideal}

\begin{figure}
\null\vskip 1cm
\includegraphics[width=8cm]{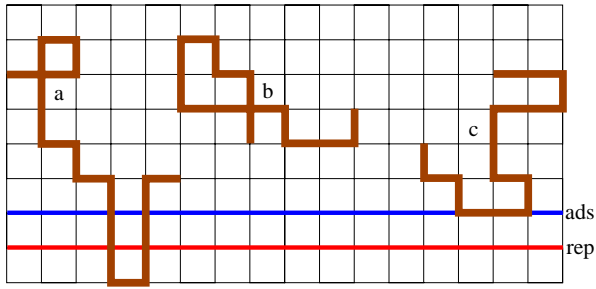}
\caption{
\label{fig:boundarycond}
(Color online) Ideal polymers on a ($d=2$) square lattice close
to a repulsive boundary (red horizontal line denoted ``rep"), possibly
covered with an adsorbing layer (blue horizontal line denoted ``ads").
Configurations that cross or touch the repulsive surface (such as
 ``a") must be discarded. Those that do not
approach either of the surfaces (such as  ``b") all have
 weight 1. In the presence of the attractive (adsorbing) layer,
configurations that touch but do not cross this layer gain an extra weight
$q={\rm e}^{-V/k_BT}$ for every point on the surface; the configuration
``c" thus has weight $q^3$.
}
\end{figure}

The absence of interactions between non-adjacent monomers of an ideal
polymer significantly simplifies its treatment. For an $N$-step
polymer on a regular lattice with lattice constant $a$, such as the
square lattice depicted in Fig.~\ref{fig:boundarycond} with coordination
number $\mu=4$,  the partition function (beginning
at point $\bf r$ in free space) is simply the number of configurations
$\cal Z({\bf r},N)={\cal  N}({\bf r},N)=\mu^N$.
We will define a reduced partition function $\tilde{\cal Z}\equiv {\cal Z}/\mu^N$,
which in general should scale as $\tilde{\cal Z}\sim N^{\gamma-1}$.
In free space  $\tilde{\cal Z}=1$, and therefore $\gamma=1$.
The partition function of a polymer of $(N+1)$ steps  that begins at
$\bf r$ and ends at ${\bf r}'$ can be calculated recursively as
\begin{equation}
{\cal Z}({\bf r},{\bf r}',N+1)
=\sum_{{\bf r}''\ {{\rm nn}\ {\rm of}}\ {\bf r}'}{\cal Z}({\bf r},{\bf r}'',N),
\end{equation}
with the initial condition ${\cal Z}({\bf r},{\bf r}',0)=\delta_{{\bf r},{\bf r}'}$.
Similarly, the reduced partition function sarisfies
\begin{eqnarray}\label{eq:discretediffusion}
& \tilde{\cal Z}({\bf r},{\bf r}',N+1)-\tilde{\cal Z}({\bf r},{\bf r}',N)\nonumber \\
& =\frac{1}{\mu}\sum\limits_{{{\bf r}''}\atop{{{\rm nn}\ {\rm of}}\ {\bf r}'}}
\left[\tilde{\cal Z}({\bf r},{\bf r}'',N)-\tilde{\cal Z}({\bf r},{\bf r}',N)\right].
\end{eqnarray}
For slowly varying functions, we can employ a continuum formulation in
which the left hand side is replaced with a first derivative,
while the right hand side represents a second derivative (discrete Laplacian).
Regarding the continuous version of $N$ as a time-like variable $t$,
the continuum equation for $\tilde{\cal Z}$ becomes the diffusion equation for
the probability {\em density} $P({\bf r},{\bf r}',t)$
of a diffusing  particle that starts its motion at ${\bf r}$ and ends up at
${\bf r}'$ in time $t$, which satifies
\begin{equation}
\label{Eq:diffusionP}
\frac{\partial P({\bf r},{\bf r}',t)}{\partial t}=D{\nabla'}^2 P({\bf r},{\bf r}',t)\,,
\end{equation}
with initial condition $P({\bf r},{\bf r}',t=0)=\delta^d({\bf r}-{\bf r}')$.
The prime sign on the Laplacian indicates spatial derivatives
with respect to ${\bf r}'$.
The diffusion constant $D$ is chosen such that in free space the
mean squared distance coincides with the random walk value of
$\langle({\bf r}-{\bf r}')^2\rangle=a^2N=2dDt$ in $d$ space dimensions,
and thus $D=a^2/2d$. The probability {\em density} $P$ is related
to the discrete probability $\tilde{\cal Z}$ by $\tilde{\cal Z}=a^dP$.

In free space all configurations have identical weight. However, in the
presence of a {\em repulsive} wall,  such as depicted by the lower (red) horizontal
line in Fig.~\ref{fig:boundarycond}, walks that touch or cross that line,
 such as walk ``a" in the figure, must be eliminated from consideration.
This can be achieved by applying Eq.~\eqref{eq:discretediffusion} only to the points
$\bf r$ above the repulsive line, while setting $\tilde{\cal Z}({\bf r},{\bf r}',N)=0$,
whenever ${\bf r}'$ is on the repulsive boundary. The continuum limit for $P$ will
then correspond to the solution of Eq.~\eqref{Eq:diffusionP}
with {\em absorbing} boundary conditions.

Since the statistical weight of every path is independent of its direction,
\begin{equation}\label{eq:reciproc}
\tilde{\cal Z}({\bf r},{\bf r}',N)=\tilde{\cal Z}({\bf r}',{\bf r},N),
\end{equation}
i.e. there is symmetry with respect to interchange of the start and end points of the chain.
Consequently, in the diffusion equation~\eqref{Eq:diffusionP}, the prime can be removed from the Laplacian.
(This is the usual reciprocity relation of the diffusion problem~\cite{weiss_book}.)
After such a change, both sides of the modified Eq.~\eqref{Eq:diffusionP} can be integrated
over ${\bf r}'$, the resulting survival probability $S({\bf r},t)\equiv\int P({\bf r},{\bf r}',t){\rm d}^d{\bf r}'$ evolving as
\begin{equation}
\label{Eq:diffusion}
\frac{\partial S({\bf r},t)}{\partial t}=D\nabla^2 S({\bf r},t),
\end{equation}
with absorbing boundary conditions.  The initial condition
for survival probability is $S({\bf r},t=0)=1$, everywhere inside the
space where the particle can diffuse, and $S({\bf r},t)=0$ on the
absorbing boundaries. This quantity coincides with the total reduced
partition function $\tilde{\cal Z}({\bf r},N)$.
Our previous works considered a variety of cases with scale-free
repulsive boundaries~\cite{MKK_EPL96,MKK_PRE86,HK_PRE89,HK_JCP141,AK_PRE91,HK_PRE92},
while in this work we are mostly interested in attractive boundaries
or in mixtures of the two.

When a repulsive surface is covered by an attractive (adsorbing) layer
(blue top horizontal line in Fig.~\ref{fig:boundarycond}), every
time a polymer visits that layer its statistical weight is increased
by a factor $q={\rm e}^{-\beta V}$, where $V<0$ is the energy gain.
(For $q=1$ the layer has no effect, while $q<1$ corresponds to a repulsive potential.)
The partition function  can now be calculated from
\begin{equation}\label{eq:iterate}
{\cal Z}({\bf r},{\bf r}',N+1)
=x({\bf r})\sum_{{\bf r}''\ {{\rm nn}\ {\rm of}}\ {\bf r}'}{\cal Z}({\bf r},{\bf r}'',N),
\end{equation}
where $x({\bf r})=q$, for $\bf r$ in the adsorbing layer, and $x({\bf r})=1$, otherwise.
This equation must be supplemented with the initial condition
${\cal Z}({\bf r},{\bf r}',0)=x({\bf r})\delta_{{\bf r},{\bf r}'}$ to ensure reciprocity.
The usual diffusion equation, still applicable outside the absorbing layer,
is thus modified by the potential near the  layer.
It is important to note that the region below the absorbing layer is still
impenetrable, and thus the partition function is strictly zero below the surface.

The phenomenology of polymer absorption is as follows:
For zero temperature ($q\to\infty$) all monomers are on the absorbing layer,
and the partition function is dominated by the energy contribution.
At small, but finite, temperatures parts of the polymer detach from the
surface gaining entropy. (We can assume that one end of the polymer is
always attached to the surface to avoid discussion of the center of mass entropy.)
The average number of visits to the absorbing layer will be proportional to $N$ ($n=cN$,
with $c$ depending on temperature).
Despite the loss of entropy, the energy gain from such visits leads to a partition function
${\cal Z}_a\simeq [z_a(T)]^N\gg \mu^N$ in the  {\em adsorbed} phase.
As temperature increases and $q$ is reduced, there is a  point where the
decrease in the number of configurations due to the impenetrable boundary is exactly
compensated by the extra weight provided by the adsorbing layer to configurations that
touch the adsorbing layer. From the perspective of a random walker,
the reduction in the number of possible paths by the boundary is exactly
made up by the extra weight of the walks that arrive at the attractive potential (blue line)
but do not touch the absorbing boundary (red line).

The adsorption of a {\em discrete} ideal polymer was studied by Rubin~\cite{Rubin65,Rubin84}.
He determined the transition point $q_c$, and demonstrated that for a planar attractive surface
${\cal Z}=\mu^N$, i.e. $\gamma_a=1$.  Comparing the behavior of an ideal polymer at $T_a$,
to a random walk (diffusion) with {\em reflecting} boundary conditions, Rubin concluded that
for large $N$ these two problems coincide, although subtle differences remain for small $N$.
Clearly, in the presence of reflecting boundaries the survival probability of a diffusing particle is
always $S({\bf r},t)=1$, which corresponds to a polymer at the  adsorption
transition point with $\tilde{\cal Z}({\bf r},N)=1$.

The universal aspects of Rubin's results can be captured in the continuum limit,
taking advantage of the mapping between configurations of the ideal random walks (path integral),
and quantum mechanics of a particle in a potential~\cite{Gennes69}.
In particular the (ideal) polymer adsorption problem is mapped to a quantum particle in a
one-dimensional potential of an attractive well adjacent to an impenetrable barrier.
Depending on the strength of attraction, such a potential may or may not admit a bound state.
The bound state (corresponding to the absorbed polymer) has a wave-function
decaying as $\psi(z)\sim e^{-\lambda z}$ away from the potential; its energy ($\propto -\lambda^2$)
designating the gain in polymer free energy on adsorption. As the potential is weakened,
$\lambda$ vanishes (linearly in $T_a-T$) indicative of the adsorption transition
point. At coarse-grained level, the combination of barrier and potential can be
expressed as the mixed (Robin) boundary condition $\psi'+\lambda\psi=0$.
Under further coarse-graining, at scales larger than $\lambda^{-1}$ (irrespective of its sign),
this requirement becomes equivalent to the Dirichlet boundary condition $\psi=0$,
while for $\lambda=0$  (an unstable fixed point under coarse-graining), it is the Neumann
boundary condition $\psi'=0$.
From the perspective of random walks, $\psi=0$ corresponds to absorbing boundaries,
and $\psi'=0$ to reflecting boundaries; both limits are scale invariant (i.e., such boundaries
do not introduce a new length scale to the polymer problem.)

The above considerations lead to the following interesting result:
If {\em all} the confining boundaries and inclusions immersed in a long ideal
polymer are at adsorption transition point, and thus in the corresponding diffusion problem
all barriers are reflective, then the trivial solution of Eq.~\eqref{Eq:diffusion}
is $S({\bf r},t)=1$ for any $t$. As this does not depend on the positions of
the various obstacles, there can be no polymer-mediated force between them!
Note that this is true for arbitrary shapes, and the boundaries do not need to be scale-free.
(In the particular case of scale-free surfaces, we note the result $\gamma_a=1$ for future reference.)

Analytical solutions of Eqs.~(\ref{Eq:diffusionP} and \ref{Eq:diffusion})
are available for a number of simple shapes~\cite{Carslaw59}.
For scale-free shapes it is convenient to choose a coordinate system centered on the
center of symmetry (such as the tip of a cone). The
dimensionless survival probability $S$ can only depend on the dimensionless vector
${\bf w}={\bf r}/\sqrt{Dt}$. Thus, $S({\bf r},t)=H({\bf w})$,
and Eq.~\eqref{Eq:diffusion} reduces to
\begin{equation}
\label{Eq:diffusion_red}
\nabla_{\mathbf w}^2 H+\frac{1}{2}{\bf w}\cdot\vec{\nabla}_{\mathbf w}H=0,
\end{equation}
where the subscript ${\bf w}$ indicates derivatives with respect to components of
${\bf w}$. In terms of these dimensionless variables, either the function $H$
or its normal derivative vanish on the absorbing or reflecting surfaces
respectively.  For some geometries, the solution to Eq.~\eqref{Eq:diffusion_red}
can be expressed in terms of a radial distance $w$, and a combination of angular
variables, such as the polar angle $\theta$ and $(d-2)$ azimuthal angles
$(\phi, \psi, \cdots)$. For $w\ll 1$, i.e. for long times $t$,  the distance
dependence is expected to be a simple power law $w^{\eta}\Psi(\theta,\phi,\dots)$.
In this limit, the second term in Eq.~\eqref{Eq:diffusion_red} becomes negligible,
and the problem reduces to solving the Laplace equation
\begin{equation}\label{eq:Laplace}
\nabla_{\mathbf w}^2 (w^{\eta}\Psi)=0.
\end{equation}
For small fixed {\bf r} we have $S\sim t^{-\eta/2}$, and comparing it with the expected
$\tilde{\cal Z}\sim N^{\gamma-1}$, we find that $\eta=2(1-\gamma)$,
i.e. it is the same exponent $\eta$ that appears in Eq.~\eqref{Eq:gammanu} for any polymer type.

Thus obtaining the exponent $\gamma$,
and the related force amplitude, is reduced to finding $\eta$ in the solution of
Eq.~\eqref{eq:Laplace} with appropriate boundary conditions. If all boundaries
are attractive, then we already know the solution, corresponding to $\gamma_a=1$.
If all  surfaces are repulsive,then  the solution to absorbing conditions of the diffusion
equation will need to vanish on the boundaries.
Problems of this type have been solved for a variety of geometries in the past~\cite{BenNaim,MKK_EPL96,MKK_PRE86,HK_PRE89,HK_JCP141,AK_PRE91}.
In the following Sections we consider several examples of mixed boundaries.

\begin{figure}
\null\vskip 1cm
\includegraphics[width=5cm]{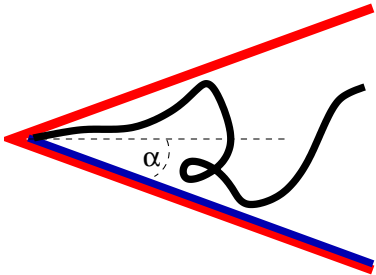}
\caption{
\label{fig:Two-sided_wedge}
(Color online) A two-dimensional wedge of inner angle
$2\alpha$, with a polymer anchored close to  its tip. The top side of the
wedge is repulsive, while the bottom side is at adsorption transition point.
}
\end{figure}

\section{Two-dimensional ideal polymers near mixed boundaries}\label{sec:mixed2D}

In a $d=2$ scale-free geometry the (Laplace) Eq.~\eqref{eq:Laplace}
simplifies  to $\Psi''+\eta^2\Psi=0$, where prime denotes the
derivative of $\Psi$ with respect to the angle $\theta$. This equation
is solved by  linear combinations of $\sin(\eta\theta)$ and
$\cos(\eta\theta)$. For repulsive or attractive boundaries of the
polymer problem, we must use boundary conditions of vanishing $\Psi$
or vanishing derivative $\Psi'$, respectively. This should be viewed as an
eigenvalue equation, and the primary goal is finding the correct
value of $\eta$. This equation (complemented by the boundary
conditions) has many eigensolutions. Since $\Psi$ determines the probability or partition
function, its sign cannot change, and consequently we are
interested in the ``ground state" that corresponds to the lowest
value of $\eta$.

\begin{figure}
\includegraphics[width=7cm]{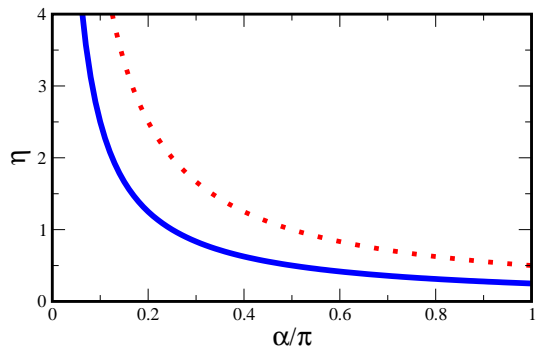}
\caption{
\label{fig:Eta2DWedge}
(Color online) The solid line depicts the exponent $\eta$ for a mixed
repulsive/attractive wedge as depicted in Fig.~\ref{fig:Two-sided_wedge}, with opening angle
$2\alpha$. For comparison, the dotted line depicts $\eta$
for a wedge with two repulsive boundaries.
}
\end{figure}

As an example, consider an ideal polymer anchored close to the central
angle $2\alpha$ of a two dimensional wedge with mixed (repulsive/attractive)
boundaries as depicted  in Fig.~\ref{fig:Two-sided_wedge}.
If the angle $\theta'$ is measured from the lower boundary, then the appropriate
solution is $\Psi=\cos(\pi\theta'/4\alpha)$, i.e.,
\begin{equation}
\label{Eq:eta_mixed_wedge}
\eta=\frac{\pi}{4\alpha}\quad{\rm or}\quad \gamma=1-\frac{\pi}{8\alpha},
\end{equation}
For  {\em two} repulsive boundaries the eigenfunction is $\Psi=\sin(\pi\theta'/2\alpha)$,
corresponding to
\begin{equation}
\label{Eq:eta_repulsive_wedge}
\eta=\frac{\pi}{2\alpha}\quad{\rm or}\quad \gamma=1-\frac{\pi}{4\alpha}.
\end{equation}
Figure~\ref{fig:Eta2DWedge} depicts the $\alpha$-dependence of $\eta$
both for mixed and for purely repulsive boundaries.
Note, that in both cases the exponent diverges  in the limit of $\alpha\to0$,
capturing the vanishing of the available states.
Equation~\eqref{Eq:eta_mixed_wedge} with $2\alpha=\pi$, is the same as
Eq.~\eqref{Eq:eta_repulsive_wedge} for $2\alpha=2\pi$, both representing a single
repulsive semi-infinite line with $\eta=1/2$. This means that
in $d=2$ the presence of such obstruction does limit the behavior
of an ideal polymer. (An obstructed line only has marginal effects in $d=3$.)
It is interesting to note that while the mixed wedge is equivalent to a repulsive wedge
of twice the angle, it can explore configurations not accessible to the repulsive case
with $2\alpha>2\pi$, i.e., beyond what would be possible in $d=2$.

The above results for $d=2$ are also applicable to wedges in any dimension $d$,
since the function $\Psi$  is independent of coordinates parallel to the edge.
In particular, in $d=3$ the $2\alpha=\pi$ case
of Eq.~\eqref{Eq:eta_mixed_wedge} corresponds to a plane half of which
is repulsive while the other half is attractive, with $\gamma=3/4$.

\begin{figure}
\includegraphics[width=6cm]{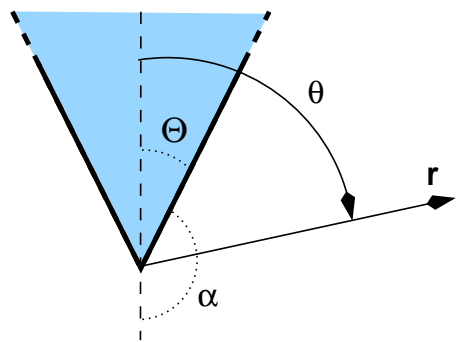}
\caption{
\label{fig:Coordinates}
(Color online) Coordinates for wedge (in $d=2$) or circular cone (in $d=3$).
The apex angle of a cone or wedge that excludes the polymer is denoted by
$\Theta$, and its compliment by $\alpha=\pi-\Theta$. The coordinate angle
$\theta$ will be measured from the axis of the cone or wedge.
}
\end{figure}

The above expressions for $\eta$ enable us to compute the force amplitude $\cal A$
in the situation depicted in Fig.~\ref{fig:def_geometry} in $d=2$, when
a wedge with a polymer attached to it approaches an excluded half space.
We simply need to find the exponents in the two extreme situations,
when the wedge is either far away, or is touching the line.
Anticipating the generalization from the wedge in $d=2$ to the cone of opening angle
$\Theta$ in $d=3$ in the next section, we shall use the notation depicted in Fig.~\ref{fig:Coordinates}.
The isolated cone in Fig.~\ref{fig:extremes}(a) has purely repulsive boundaries and is solved by
Eq.~\eqref{Eq:eta_repulsive_wedge} giving $\eta_{\rm far}=\pi/[2(\pi-\Theta)]$, while the
touching of cone and plate in Fig.~\ref{fig:extremes}(b) is described by mixed boundary
situation and Eq.~\eqref{Eq:eta_mixed_wedge}
results in $\eta_{\rm near}=\pi/(\pi-2\Theta)$. (We have set $\alpha\to\pi-\Theta$ for the repulsive wedge,
and $\alpha\to\pi/4-\Theta/2$ in the mixed case.) This results in the force amplitude
\begin{equation}
{\cal A}=\frac{\pi^2}{2(\pi-2\Theta)(\pi-\Theta)}.
\end{equation}
Note that $\cal A$ is always positive, even in the limit of a ``needlelike" wedge  with $\Theta\to0$.
In the reversed situation where the cone is attractive, while the plane
is repulsive, $\eta=0$ in the remote configurations but retains the same value as
before when the plane and cone are in contact, leading to
\begin{equation}
{\cal A}=\frac{\pi}{\pi-2\Theta}.
\end{equation}
This amplitude is larger than in the previous example,
and in the $\Theta\to0$ limit is the same as a polymer
that is brought to the vicinity of a repulsive surface while held
at an endpoint  {\em without a wedge}.

\begin{figure}
\includegraphics[width=8cm]{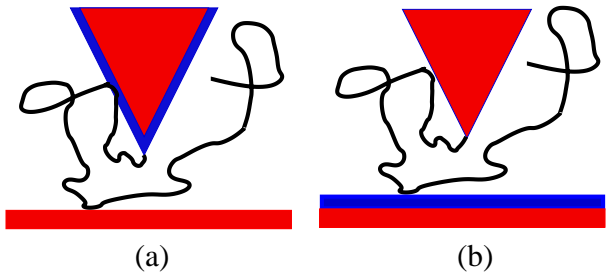}
\caption{
\label{fig:two_mixed_systems}
(Color online) Two variants of cone-plane polymer-mediated setups with mixed boundary
conditions: In $d=3$ we can have (a) an attractive cone and repulsive
plane, or (b) repulsive cone and attractive plane. In
$d=2$ these drawing can be viewed as a wedge approaching a line.
}
\end{figure}

\section{Three-dimensional ideal polymers near mixed boundaries}\label{sec:mixed3D}

In the presence of azimuthal symmetry in three dimensions,
as  with cones of circular cross section, or such a cone touching
a plane and perpendicular to it, the Laplace Eq.~\eqref{Eq:diffusion_red} simplifies.
With $\Psi$ depending only on the polar angle $\theta$ as illustrated in Fig.~\ref{fig:Coordinates},
it takes the form
\begin{equation}
\label{Eq:Legendre3D}
(1-u^2)\frac{d^2\Psi}{du^2}-2\mu\,\frac{d\Psi}{du}
+\eta(\eta+1)\Psi=0,
\end{equation}
where $u\equiv \cos\theta$.
We seek a regular eigensolution $\Psi(\theta)$
that vanishes on repulsive boundaries or has a vanishing derivative on attractive boundaries.
The general solution to this equation is given by regular (rather than associated) Legendre functions
\begin{equation}
\label{Eq:3D_solution}
\Psi(\theta)=a_1P_{\eta}(u)+a_2Q_{\eta}(u).
\end{equation}
Note that $P_\eta(1)=1$ for any $\eta$, while $P_\eta(-1)$ diverges for noninteger $\eta$.
Similarly, $Q_\eta(\pm1)$ is divergent. The linear combination in Eq.~\eqref{Eq:3D_solution}
can be made regular at $-1$ by a proper choice of $a_1/a_2$. In
Refs.~\cite{MKK_EPL96,MKK_PRE86,HK_PRE89} we described the analytical
solutions of this equation for purely repulsive cones, or such cones
touching a repulsive plane. (Geometries without azimuthal symmetry can be easily
handled numerically~\cite{AK_PRE91}.)

In many cases, the regularization procedure can be avoided by a
convenient choice of functions. For the geometry depicted in
Fig.~\ref{fig:Coordinates},  the solution must be regular for
$\Theta\le\theta\le\pi$. Instead of using combinations of $P_\eta$
and $Q_\eta$, we can simply use $P_{\eta}(-\cos\theta)$, which will
be regular at $\cos\theta=-1$. The value of $\eta$ for a {\em repulsive}
cone is then determined by requiring
\begin{equation}
\label{Eq:eigen_cone}
P_{\eta}(-\cos\Theta)=0.
\end{equation}
Since $\Psi$ cannot change sign in the physically permitted region, the
smallest possible $\eta$ must be chosen. This procedure is described in detail
in Refs.~\cite{MKK_EPL96,MKK_PRE86,HK_PRE89}.  An attractive boundary
requires $\partial\Psi/\partial\theta$ to vanish at $\theta=\Theta$. This,
as in all  cases of purely attractive boundaries, is trivially achieved by
$\Psi={\rm constant}$, and $\eta=0$.

\begin{figure}
\includegraphics[width=8cm]{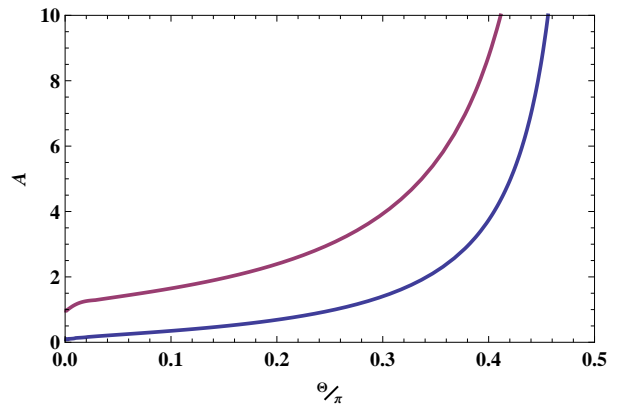}
\caption{
\label{fig:forceconst}
(Color online) The force amplitude for a three-dimensional
cone with apex angle $\Theta$ approaching a plane. The top curve corresponds to an
attractive cone and repulsive plane, while the bottom curve is for a repulsive cone  and
attractive plane.}
\end{figure}

Calculation of $\eta$ for the situation when the cone touches a plane
(as in Fig.~\ref{fig:extremes}b) is simplified by noting
that for non-integer $\eta$, $P_\eta(u)$ and $P_\eta(-u)$ are
linearly independent and both solve Eq.~\eqref{Eq:Legendre3D}~\cite{NISTlib}.
Thus for non-integer $\eta$, Eq.~\eqref{Eq:3D_solution} can be replaced by
\begin{equation}
\label{Eq:simplified}
\Psi(\theta)=P_{\eta}(\cos\theta)\pm P_{\eta}(-\cos\theta).
\end{equation}
The $-$ and $+$ signs enforce vanishing of $\Psi$ or its derivative at
$\theta=\pi/2$, respectively, corresponding to a repulsive or attractive
plane. The remaining boundary condition at $\theta=\Theta$ can be
implemented by proper choice of the exponent $\eta(\Theta)$, which can be
obtained numerically.
New results pertain to the mixed boundary setups
depicted in Fig.~\ref{fig:two_mixed_systems}.
(The case of all attractive boundaries is trivial, while that of
repulsive boundaries was considered in Refs.~\cite{MKK_EPL96,MKK_PRE86,HK_PRE89}.)
From knowledge of the values of $\eta$ in situations
depicted in Fig.~\ref{fig:two_mixed_systems}, we can use
Eq.~\eqref{Eq:A} to determine the force amplitudes.
Figure~\ref{fig:forceconst} depicts the force amplitude for the two
cases as a function of cone apex angle $\Theta$. Similarly,
to analogous solution in $d=2$ the force amplitude for repulsive
plane and attractive cone is larger than that for the reversed situation.

\begin{figure}
\includegraphics[width=8cm]{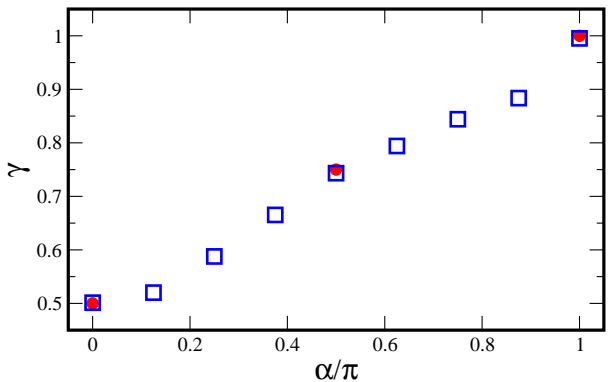}
\caption{
\label{fig:DecoPlaneGammaVsAlpha}
(Color online) Exponent $\gamma$ for an ideal polymer attached to a
repulsive surface decorated by an attractive sector with the central angle
$2\alpha$, as depicted in Fig.~\ref{fig:DecoPlane}. The adsorbing sector is
assumed to be at the adsorption transition point. Open squares depict
results of direct numerical estimation of $\gamma$, while the full circles
represent the exactly known values.
}
\end{figure}

For a less symmetric setup in $d=3$, the Laplace equation Eq.~\eqref{eq:Laplace} will
depend on two angular variables, and implementation of boundary conditions may prove difficult.
Fortunately, a direct numerical implementation of Eq.~\eqref{eq:iterate}, typically only requires
a few thousand iterations for a good estimate of $\gamma$ from the
dependence of $\tilde{\cal Z}$ on $N$. As an example
we considered a repulsive plane decorated by an attractive sector of
opening angle $2\alpha$ as depicted in Fig.~\ref{fig:DecoPlane}, and
results from a numerical estimation are shown in Fig.~\ref{fig:DecoPlaneGammaVsAlpha}.
For $\alpha=0$ we naturally recover $\gamma=1/2$, as for a purely repulsive plane,
while $\gamma=1$ for $\alpha=\pi$, corresponds to a purely attractive plane.
For intermediate angles, the critical exponent simply interpolates between
these two limiting values. At $\alpha=\pi/2$ the plane is divided into
repulsive and attractive halves, with $\gamma=3/4$ as we found in an equivalent geometry in $d=2$.

\section{Discussion}

In this work we considered several cases of polymer-mediated interactions between
repulsive or attractive surfaces. While the loss of entropy leads to a repulsive force
between impenetrable obstacles, the gain in energy may cause an attractive force
for absorbing surfaces.
If the confining objects do not introduce a length scale, which is the case for impenetrable
obstacles, and surfaces at the adsorption transition point, the entropic force is dimensionally
constrained to the form ${\cal A}\,k_BT/h$ where $h$ is a characteristic distance between
the scale-free surfaces. The amplitude $\cal A$ depends on geometry and universality class
of the polymer system (ideal, self-avoiding, or $\theta$ polymer).

A hypothetical setup, such as in Fig.~\ref{fig:two_mixed_systems}, involves a long polymer
(or several polymers) of length $N a$ attached to the tip of a cone at a separation $h$ from a plane.
For an adsorbing surface before the desorption transition, a typical configuration will
consist of a nearly straight segment of the polymer stretched from the cone tip to the surface,
followed by a much longer segment absorbed to the surface.
This situation will hold as long as $Na\gg h$ and $Na \gg \xi$, where $\xi$
is a characteristic segment size that diverges close to adsorption transition
point as $\xi\propto (T-T_a)^{-1/\phi}$~\cite{Eisenriegler82,debell}.
The free energy difference between adsorbed and free polymers vanishes at the
adsorption transition point as $\Delta F_a\sim -Nk_BT(a/\xi)$. The adsorbed
polymer also fluctuates away from the surface, forming a layer of thickness
$\ell\propto \xi^\nu$ that also diverges at the adsorption transition point.
Equation~\eqref{Eq:force} should apply only in the separation range $\xi\gg h\gg \ell$,
where the short and long-scale cutoffs are immaterial.
On shorter scales, the force should saturate, presumably to order of $k_BT/\ell$,
while at large scales, the polymer should be stretched, with the force reduced to
$k_BT/\xi$, corresponding to the loss of free energy per unit length.
We thus expect the following sequence of crossovers for the force
\begin{equation}
    F=
    \begin{cases}
     -\frac{k_BT}{\xi} & {\rm if}~ h\gg\xi  \\
      {\cal A}\frac{k_BT}{\xi} & {\rm if}~ \xi\gg h\gg\ell \\
      \sim\pm \frac{k_BT}{\ell} & {\rm if}~ h\ll\ell
    \end{cases}\,.
\end{equation}
(The amplitude ${\cal A}$, and the sign of the force, is determined by surface and polymer types.)
Closer still to the transition, such that $\xi\sim Na$, additional crossovers are expected
that are not discussed here.

 For separations of order of 0.1 $\mu$m at room temperature, the entropic force is of
order 0.1pN. Forces of such magnitude are now measurable by a host of single molecule
manipulation techniques~\cite{bustamante03,kellermayer,neuman,deniz,neuman08}, e.g. by
atomic force microscopes (AFMs)~\cite{fisher}, microneedles~\cite{kishino}, and
optical~\cite{neuman2004,hormeno} and magnetic~\cite{gosse} tweezers.
With a good AFM tip, distances can be measured to accuracy of a few
nanometers~\cite{kikuchi97,neuman08}, with forces of order of 1~pN
measured in nearly biological conditions~\cite{brk1997,drake89}.
We note that these accuracies fall within the range of entropic forces for
fluctuating, featureless polymers described above.

While we considered here the case of a single polymer, related fluctuation-induced
forces are also expected in the case of dense melts of long polymers. Such forces
have been proposed (dubbed anti-Casimir forces) for dense polymer melts between
parallel plates~\cite{Semenov05,Obukhov05}. For the scale free geometries that we
propose, these forces should have the general forms proposed in this paper,
albeit with different universal amplitudes.

Finally, we note the interesting observation about the lack of  polymer mediated
forces for any number of objects (scale-free or not) immersed in {\em ideal} polymers,
as long as all surfaces are at the special adsorption transition point.
Compensation of  the loss of entropy by marginally attractive energies renders
such obstacles invisible to ideal polymers, in a situation similar to index
matching of colloids by a fluid of the same dielectric constant.
(The van der Waals interaction vanishes in such a case.)
It is tempting to imagine that such a situation can also occur for objects
in a self-avoiding polymer. However, the results ($\gamma_a>\gamma_1$) in Table I indicate
that ${\cal A}<0$ for a self-avoiding polymer near a plane at adsorption transition point.
Nonetheless, by appropriate coatings of the surfaces (as in Fig.~\ref{fig:DecoPlane}) it should
be possible to reduce  the force prefactor to ${\cal A}=0$. Thus there is indeed hope
for engineering (at least scale-free) obstacles that are force-free in
a self-avoiding polymer solution.

\begin{acknowledgments}
This work was supported by the National Science Foundation under Grant
No.~DMR-1206323 (M.K.), and
the Israel Science Foundation Grant No.~453/17 (Y.K.).
\end{acknowledgments}
\end{document}